# From Reconnaissance to Response: Quantitative Risk Parameterization and Game Theoretic Containment in Modern Enterprise Attack

Shadeeb Hossain [ORCID ID: 0000-0002-5224-7684], Elias Calboreanu [0009-0008-9194-0589]

***Abstract*— Modern Security Operations Centers (SOCs) struggle with delayed manual incident response, enabling adversaries to advance through the Cyber Kill Chain (CKC) during early-stage reconnaissance. While classical game-theoretic defense models optimize strategic resource allocation, they rely on static utility matrices that fail to adapt to dynamic telemetry. This paper presents an integrated, metrics driven decision engine that bridges quantitative risk parameterization and continuous automated response time. Common Vulnerability Scoring Systems (CVSS v3.1) exploitability parameters are mapped to attacker's success probabilities and evaluate defender's log distributions via Factor Analysis of Information Risk (FAIR) Monte Carlo simulations. Real-time SIEM logs streams are modeled as Poisson process arrival rates ($\lambda$), dynamically updating defender's posterior threat belief $\mu(t)$ through sequential Bayesian filtering. A closed form threshold ($\mu^*$) is derived by framing the interaction as a dynamic Bayesian-Stackelberg game, where the expected unmitigated risk exceeds proactive containment cost ($C_D$). Parameterized against empirical data from the 2023 MGM Resorts and Caesars Entertainment cyber incident ($C_D = \$15M, E[L_{FAIR}] = \$92.58M, P(E) = 0.739$), simulation results demonstrate that the engine suppresses transient background noise while triggering automated SOAR network isolation within seconds of adversarial probing. Multi-parameter sensitivity analysis confirms that the decision boundary dynamically adjusts to live perimeter vulnerability, offering a control-theoretic foundation for sub-minute automated threat containment.**

## I. INTRODUCTION

The Cyber Kill Chain (CKC) framework enables the systemic dissection of a complex, multi-stage cyber threat into a sequence of seven distinct phases: (i) *Reconnaissance*, (ii) *Weaponize*, (iii) *Delivery*, (iv) *Exploitation*, (v) *Installation*, (vi) *Command and Control (C2)*, and (vii) *Act on Objective* [1,2]. By mapping the adversary's progression through these sequential operational milestones, the framework provides a structured foundation for defenders to optimize defensive posture, detect anomalies early, and allocate mitigation resources efficiently to prevent lateral penetration [3,4].

As the foundational phase of the CKC, Reconnaissance dictates the ultimate trajectory of a cyber-attack [5]. During this phase, adversaries passively or actively probe target infrastructures to harvest intelligence regarding active system applications, network topologies, and unpatched software vulnerabilities [5-7]. Early-stage reconnaissance techniques include OS fingerprinting, social engineering, and public metadata analysis; however, detecting an adversary at this boundary presents a significant challenge [8]. Disrupting the kill chain at this phase yields the highest strategic utility for the defender, as it denies the attacker the critical intelligence required to engineer a successful weaponized payload.

To model this highly dynamic adversarial interaction mathematically, recent research has increasingly turned to Game Theory [9, 10]. Game theory models the strategic interactions between rational agents, namely defenders and attackers; enabling the formal analysis of complex decisions in interactive setting [11]. By quantifying the strategic choices available to each player, game theory solves for operational states such as Nash Equilibrium, where the attacker and defender can neither improve their payoff by altering their strategy [12]. These models allow organizations to transition from reactive patching to proactive dynamic resource allocation based on mathematically proven optimal mixed strategies.

A major limitation of conventional game-theoretic security models lies in their relevance on static, predefined utility matrices. However, in real world cybersecurity network, static payoff values fail to capture the dynamic nature of adaptive adversaries, that continuously adjust their tactics, techniques and procedures (TTP) based on feedback during reconnaissance phase [4].

To bridge this gap, modern models will require empirical parameterization, wherein player utilities are continuously calibrated using live security telemetry. By being able to map

Shadeeb Hossain is affiliated as the Founder and Principal Engineer at Research Division, Shadeeb Engineering Lab, Brooklyn, NY 11223, USA [e-mail: shadeeb@shadeebengineeringlab.com]. He is also appointed as an Adjunct Faculty member and Dissertation Chair at Capitol Technology University, *Department of Engineering, Laurel, MD 20708 [e-mail: shossain@captechu.edu]*.

Elias Calboreanu is a PhD Candidate at Capitol Technology University. [e-mail : ecalboreanu@captechu.edu].

the attacker rewards to standardized Common Vulnerability Scoring System (CVSS) [13] exploitability metrics and deriving defender losses from empirical asset valuation frameworks, such as Factor Analysis of Information Risk (FAIR) [14-16] and Intrusion Detection System (IDS) [17,18]; game theoretic models can dynamically update payoff functions. This transition from static metrics to empirical data driven parameterization ensures that calculated defense strategies remain optimal against evolving adaptive threat vectors across multi-tier network architectures.

This study conducts a comparative analysis of modern attack paths to demonstrate how to integrate standard security metrics; specifically, CVSS exploitability parameters, FAIR framework quantitative loss metrics, and dynamic SIEM telemetry, into an automated decision-making framework that yields superior, mathematically grounded response strategies compared to manual analysis.

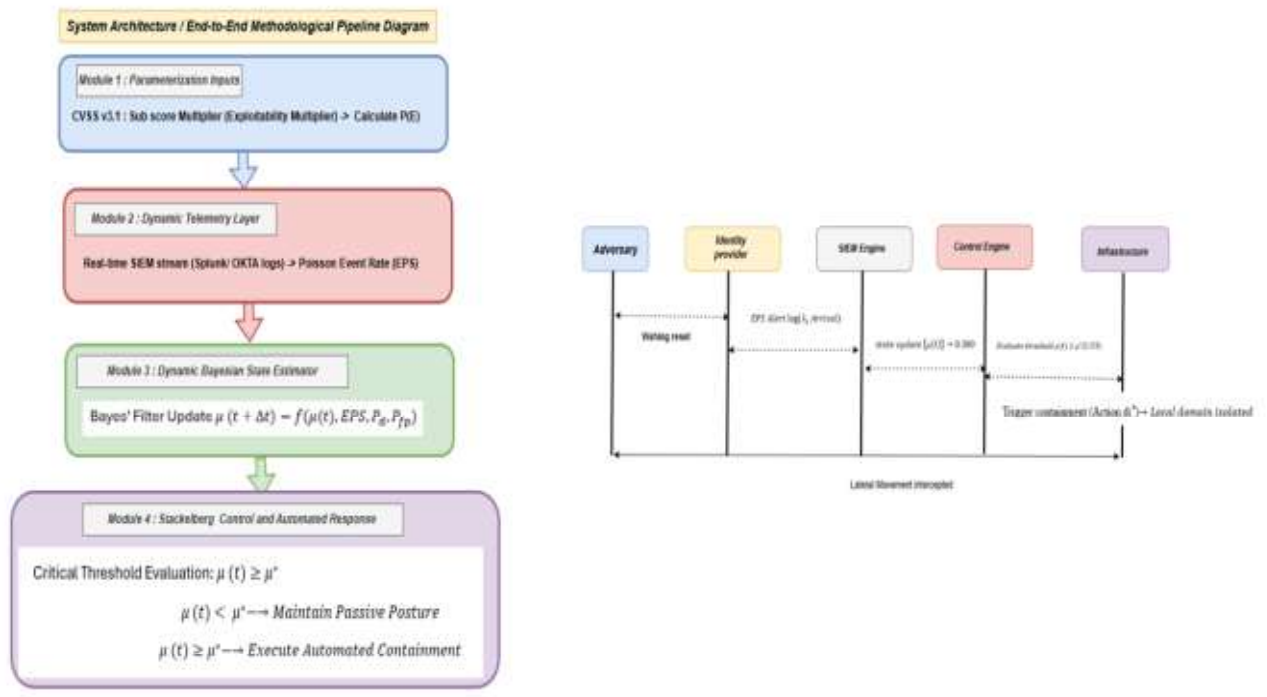

**Fig. 1:** System architecture of the methodological pipeline**.**

Fig.1 shows the system architecture with the end-to-end mathematical parameterization and automated control pipeline. Static vulnerability metrics (CVSS v 3.1) and FAIR Monte Carlo financial loss distribution [$(E[L_{FAIR}]$] calibrates the game-theoretic payoff matrix. Real-time SIEM log streams (EPS) update the defender's Bayesian belief state ($\mu(t)$), triggering an automated containment the instant the threat probability crosses the optimal Stackelberg decision threshold.

The remainder of this paper is organized into the core sections:

*(i) Related Work and Main Contributions:* Reviews foundational literature across qualitative security postures, game-theoretic signaling models and moving target defense strategies, highlighting critical gaps in empirical payoff parameterization and real-time telemetry integration.

*(ii)Background and Parameterization Framework:* Establishes the mathematical mapping between static vulnerability scores, dynamic detection probabilities, and business-impact loss functions. It also evaluates real world enterprise breaches by contrasting actual, historical defender outcomes against optimal game-theoretic countermeasures under severity ties.

*(iii) Strategic Game Formulation and Payoff Matrix:* This section discusses the optimal decision-making strategy during a cyber breach using the Game Theory approach. The strategic game matrix utility functions for the defender and adversary are discussed under two scenarios: (i) *Proactive Containment* and (ii) *Passive Reactive*.

*(iv) Simulation Results and Discussion*: This section presents numerical simulation results and sensitivity analysis demonstrating that the dynamic Bayesian -Stackelberg decision engine triggers sub-minute automated SOAR containment upon detecting sustained SIEM anomaly streams.

*(v) Limitations, Future Directions and Conclusion:* Outlines key modeling boundaries, establishes future directions for multi-state belief modeling, and concludes with strategic implications for automated SOC response engine.

## II. Related Work and Main Contributions

The mathematical modeling of cyber defense during early-stage reconnaissance has evolved from qualitative operational frameworks to dynamic-game theoretic optimizations. However, existing approaches exhibit critical limitations in empirical parameterizations and automated response execution.

### A. *Qualitative and Architectural Defensive Strategies*

Early research on enterprise cybersecurity focused primarily on qualitative evaluations of defensive postures. Ahmad *et al.* (2014) conducted a comprehensive field study analyzing organizational security strategies across operational domains, including: (i) *prevention, (ii) deterrence, (iii) surveillance, (iv) detection, (v) response, and (vi) deception* [19]. Their empirical findings revealed that while organizations universally rely on perimeter-based preventive controls, advanced strategies such as proactive deception (e.g. honeypot architecture) were virtually unused due to operational complexity. Crucially, their framework evaluated security postures qualitatively, failing to provide quantitative metrics for multi-layered defense in dynamic, high intensity attack scenarios.

### B. *Game Theoretic Modeling and Cyber Deception*

To address the dynamic nature of adversary-defender interactions during reconnaissance, recent literature has increasingly adopted game-theoretic formulations. Carroll and Grosu pioneered the application of signaling games to model network deception, solving for perfect Bayesian Equilibria under varying honeypot allocations (e.g. 10% and 15% network deception) [20]. Their model demonstrated how defenders can manipulate adversary beliefs to maximize expected payoffs during early probing. Extending deception dynamics to human behavioral variants, Schlenker *et al*., introduced the Cyber Deception Game, which optimizes defense allocations against both fully rational adversaries and boundedly rational (naïve) attackers [21].

Despite their theoretical contributions, these signaling game models suffer from two major operational drawbacks:

1. *Arbitrary Utility Matrices*: Payoff values ($U_D, U_A$) are assigned via abstract normalized constants, lacking direct connection to standardized vulnerability scores (for example, CVSS) or financial loss distribution (for example, FAIR).
2. *Static Telemetry:* Threat updating relies on static, offline probabilities rather than continuous, real-time

security information and event management (SIEM) log streams.

### C. *Moving Target Defense and Automated Control*

To counter adaptive reconnaissance in modern industrial environments, Zang *et al.* developed a moving target defense (MTD) framework for Industry 5.0 using a defender led-Stackelberg game [10]. Their approach optimizes surface-shuffling algorithms to minimize defense expenditure while preserving system availability. Similarly, Attiah *et al.* stratified defense strategies into multi-level activity tiers to evaluate multi-stage attack escalation [22]. However, these formulations treat operational containment costs as abstract variables and lack sub-minute integration with automated *Security Orchestration, Automation, and Response (SOAR)* engines.

### D. *Research Gap and Literature Comparison*

As summarized in Table-I, existing security game frameworks either rely on theoretical, uncalibrated utility parameters or lack continuous integration with live enterprise telemetry. A critical gap remains for an end-to-end framework that dynamically parameterizes player utilizes using empirical vulnerability metrics (CVSS 3.1) and financial loss distribution (FAIR Monte Carlo simulation), while continuously evaluating Bayesian threat beliefs against live SIEM Poisson event streams (EPS) to execute automated SOAR containment.

TABLE I

QUALITATIVE COMPARISON OF PROPOSED FRAMEWORK AGAINST STATE-OF-THE-ART SECURITY GAME MODELS.

| **Reference Study** | **Game model** | **Telemetry integration** | **Utility parameterization** | **Financial risk** | **Empirical calibration** | **Automated SOAR execution** |
|---|---|---|---|---|---|---|
| Ahmad et al. [19] | Qualitative survey | Statice/survey | N/A | Qualitative | Field study | No |
| Carroll and Grosu [20] | Signaling game | Static/offline | Arbitrary payoffs | None | Synthetic | No |
| Schlenker et al. [21] | Deception game | Offline probing | Pre-defined constants | None | Synthetic | No |
| Attiah et al. [22] | Multi-stage game | Discrete alert state | Static metrics | None | Synthetic | No |
| Zang et al. [10] | Stackelberg MTD | Offline logs | Heuristic costs | Operational overhead | Partial | No |
| Proposed model | Bayesian Stackelberg | Dynamic Poisson streams (EPS) | CVSS 3.1 sub-scores | FAIR Monte Carlo simulation to calculate $E[L_{FAIR}]$ | Yes (MGM/Caesars data) | Yes |

### E. *Main Contributions*

To address these limitations, this paper advances the game-theoretic cyber defense beyond static, arbitrary payoff matrices by grounding defender utilities directly in empirical vulnerability and risk parameters. The principal contributions of this work are summarized as follows:

1. *Empirical Utility Parameterization Framework*: A closed form mathematical bridge is established that maps static CVSS v 3.1 exploitability sub-scores and PERT bounded FAIR Monte Carlo financial loss distributions directly into dynamic game-theoretic utility matrices, eliminating the reliance on arbitrary payoff matrices.
2. *Continuous time Bayesian State estimation:* SIEM event ingestion streams (EPS) are modeled in real-time Poisson arrival processes, deriving sequential Bayesian likelihood updates to continuously track threat probability$\mu(t)$ against the baseline operational noise.
3. *Closed -form Stackelberg Control and SOAR Automation:* An analytical decision boundary ($\mu^*$) that dynamically balances proactive operational containment costs ($C_D$) against live perimeter risk, enabling sub-minute automated network isolation upon threat detection.
4. *Real world breach validation and sensitivity analysis :*The control engine is calibrated and validated using empirical operational data from the 2023 MGM Resorts and Caesars Entertainment cyber incidents, demonstrating robust performance across varying threat exploitability levels and noise floors.

## III. BACKGROUND AND PARAMETERIZATION FRAMEWORK

### A. *Case Studies: MGM Resorts and Caesars Entertainment Breaches*

To evaluate the effectiveness of an empirically parameterized, metrics-driven decision framework during the initial reconnaissance stage of the CKC, this research examines the 2023 cyber incidents targeting MGM Resorts and Caesars Entertainment [23,24]. In September 2023, MGM Resorts suffered a widespread cyber-attack initiated by the threat actor group Scattered Spider (an affiliate of ALPHV/BlackCat). The attack combined OSINT reconnaissance (gathering employee details using LinkedIn) with voice phishing (vishing) directed at the internal IT helpdesk to execute social engineering and bypass Multifactor Authentication (MFA). Once initial access was secured, the adversary performed internal lateral movement, compromising Okta and Azure identity provider (IdP) environments. In contrast, Caesars Entertainment was breached via a compromised third-party IT vendor endpoint using similar social engineering vectors.

The operational outcomes and defender response strategies differed drastically between the two organizations:

*Caesars Entertainment:* chose a payoff negotiation strategy, paying an estimated $15 million ransom to mitigate system lockouts and retain operational continuity.

*MGM Resorts:* refused ransom demands and instead executed manual network severance, shutting down internal servers and core operational systems. This reactive shutdown resulted in severe business interruption across reservation systems, digital room keys, and casino slot machines, causing over $100 million in direct operational and remediation losses. In both instances, high value sensitive data includes customer Personally Identifiable Information (PII) and Social Security Numbers (SSN).

### B. *Quantitative Vulnerability Modeling via CVSS v3.1*

The Common Vulnerability Scoring System (CVSS) provides an open, standardized framework to quantify vulnerability severity [25]. The CVSS exploitability metrics quantify the adversary's initial probability of success *(P(E))* during the pre-attack reconnaissance phase of the CKC.

Although the adversary in the MGM Resort's case study ultimately targeted the Identity Provider (IdP) ecosystem, the initial social engineering interface (helpdesk interactions) required no prior system credentials or Multi-Factor Authentication (MFA) from the attacker. This sets the initial exploitability parameter to its upper bound, reflecting a highly porous perimeter state that maximizes the attacker's baseline payoff (x) prior to internal reconnaissance.

In the CVSS v3.1 the Exploitability sub score formula depends on four distinct component metrics and is given by equation (1): (i) *Attack Vector (AV)*, (ii) *Attack Complexity (AC)*, (iii) *Privilege Required* (PR) and (iv) *User Interaction (UI)* [26].

$$Exploitability\ (E_s) = (8.22)\,.AV.AC.PR.UI \qquad (1)$$

To map the sub score directly into a probabilistic parameter, $E_s$ is normalized against its maximum theoretical upper bound limit (3.887) as shown in equation (2). Since no user interaction (UI: N=0.85) and privilege (PR:N=0.85) was required, the calculated exploitability sub score was 2.87, and normalized attacker's success probability was 73.9% of initial compromise.

$$P(E) = \frac{E_s}{3.887} \quad \in [0,1] \qquad (2)$$

### C. *Factor Analysis of Information Risk (FAIR) for MGM Resorts Case Study*

Factor Analysis of Information Risk (FAIR) is a stochastic risk quantification methodology used to model financial loss exposures resulting from cyber incident Loss Event Frequency (LEF) which is governed by: (i) *Threat Event Frequency (TEF)* and (ii) *Vulnerability(V)* as shown in equation (3) and (4) [27,28].

$$LEF = TEF\ .V \qquad (3)$$

where, vulnerability (V) represents the conditional probability that a threat event results in a loss event. Vulnerability can be expressed as a logistic function comparing threat capability (T) and resistance strength (R).

$$V = \frac{1}{1+e^{(R-T)}} \qquad (4)$$

Using the normalized attack success probability P (E) $\approx 0.739$ derived from equation (2), the implied capability-resistance differential$(T-R) = \ln\big(P(E)\big)/(1-P(E)) \approx 1.04$ demonstrates that the targeted posture was highly vulnerable.

To parameterize the Loss Magnitude (LM), total financial impact is decomposed into Primary Loss (PL) (direct operational downtime and incident containment) and Secondary Loss (SL) (legal liability, regulatory fines, and reputational damage) [28]. To evaluate the expected Annual Loss Exposure (ALE), a 10,000 iterations Monte Carlo FAIR simulation was expected, calibrated against the 2023 MGM Resorts cyber incident using PERT (Program Evaluation and Review Techniques) Beta distribution as shown in Fig.2.

In the FAIR parameterization pipeline, Threat Event Frequency (TEF) from equation (3) quantifies the annual rate of adversary engagement attempts. For the MGM Resorts scenario, TEF models the frequency of high capability, voice phishing and identity impersonation attacks targeting helpdesk staff. Calibrated as a PERT distribution ($TEF_{min}$ =0.5, $TEF_{mode}$= 1.14, $TEF_{max}$= 2.0 attempts/ year). Multiplying TEF by the CVSS 3.1 Vulnerability Probability (V=P (E)=0.739) establishes an empirical Loss Event Frequency (LEF) from equation (3) to approximately 0.84 events/ year. This LEF scales the Monte Carlo Loss distribution to determine the defender's expected payoff ($U_D$= -ALE) in the Stackelberg game.

For Primary Loss, the mode was parameterized at $94 million, bounded by a minimum of $80 million and a maximum of $120 Million. For Secondary Loss, the mode was set at $15 Million, with minimum and maximum bounds of $5 million and $40 million respectively. According to PERT conventions, the expected mean $\mu$, for a variable bounded on [a, b] with mode g and shape parameter $\gamma = 4$ is defined as (5):

$$\mu = \frac{(a+\gamma g+b)}{6} \qquad (5)$$

To generate Beta distributed random variables over [a, b], shape parameters α and $\beta$ for MATLAB's '*betarnd*' function were computed using the expected mean $\mu$ :

$$\alpha = \frac{(\mu-a)}{(b-a)}\left[\frac{(\mu-a)(b-\mu)}{\sigma^2} - 1\right] \qquad (6)$$

$$\beta = \alpha\left[\frac{(b-\mu)}{(\mu-a)}\right] \qquad (7)$$

From the Monte Carlo simulation results in Fig.2, the expected mean loss E $[L_{FAIR}]$ is estimated at $92.58 million with a 95$^{th}$ percentile value at risk ($VaR_{0.95}$) of $122.03 million and standard deviation of $16.87 million. This empirical loss metric establishes the payoff parameters ($U_D = -E[L_{FAIR}]$) in the game-theoretic decision framework, enabling defending organizations to elevate whether proactive mitigation or payoff negotiations yields an optimal strategic posture.

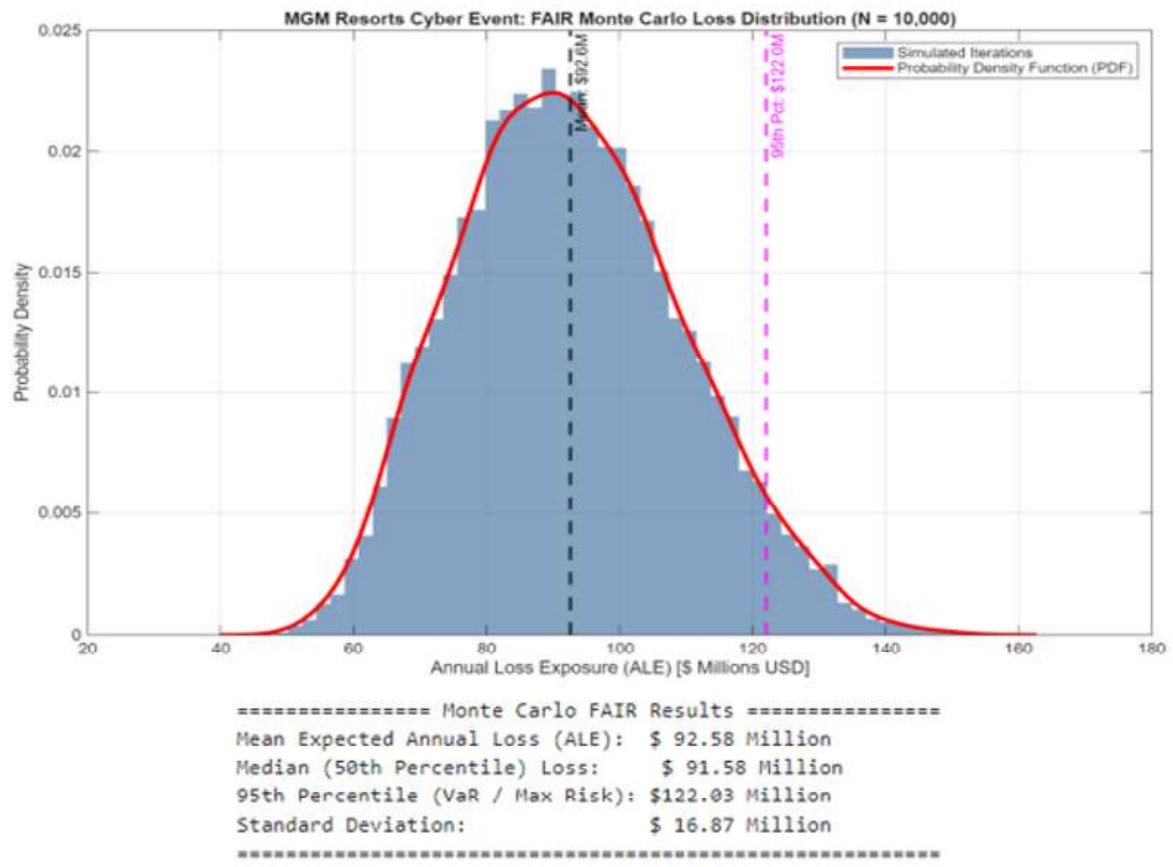

**Fig. 2:** Monte Carlo FAIR simulation of Annual Loss Exposure (ALE) for the 2023 MGM Resorts cyber incident (N=10,000 iterations).

### *A. Analyzing Security Information and Event Management (SIEM) for MGM Resorts Case Study*

Security Information and Event Management (SIEM) involves analyzing logs from various data sources to monitor and respond to potential cyber threats [29,30]. Common sources of SIEM events include: (*i) security logs detailing user logins, (ii) privilege escalations, (iii) IDS/ IPS alerts for malicious traffic patterns, (iv) blocked traffics, (v) Web server Access logs, (vi) Database Query logs, (vii) data logs from security tools, (viii) increasingly specialized logs from industrial control systems, (ix) Identity and Access Management (IAM) for authentication success or failures, (x) Logs from IaaS, PaaS and SaaS environments*.

Events per second (EPS) is a critical metric in SIEM that processes discrete login or security alerts in the network ecosystem. Mathematically it can be represented as equation (8):

$$k = \int_{t}^{t+\Delta t} EPS(\tau)d\tau \qquad (8)$$

where, k is the total number of events during a time window $\Delta t$.

*D.1 Discrete Alert State Update for a Single Signal*

Let $\theta \in \{\theta_1,\ \theta_o\}$ represent the true state of the enterprise network, where $\theta_1$ denotes an active, adversary- driven reconnaissance and $\theta_o$ represents normal operational background noise. Before observing telemetry, the defender maintains a prior attack belief $P(\theta_1 = \pi_o)$.

As SIEM collectors ingest logs across identity providers (IdPs) and endpoint detection systems, arriving alert signals z ∈ {0,1} trigger a continuous Bayesian updating process. The conditional probability of observing an alert signal =1, given an active threat ($\theta_1$) is defined by the SIEM detection sensitivity $P\langle Z = 1|\theta_1\rangle = P_d$ while the false positive rate under normal operations ($\theta_o$) is parameterized as $P\langle Z = 1|\theta_o\rangle = P_{fp}$.

Upon receiving an alert vector z, the defender updates their posterior belief $\mu\,(z) = P\langle\theta_1|z\rangle$ using Bayes' Rule (Bayesian Statistics) [31,32].

$$\mu\,(z) = \frac{P_d.\pi_o}{P_d.\pi_o + P_{fp}(1-\pi_o)} \qquad (9)$$

This dynamic posterior belief $\mu\,(z)$ reflects the real-time probability of an ongoing potential cyber-attack (or system compromise), serving as the temporal state variable that drives automated response thresholds in game-theoretic control engine.

*D.2 Continuous Time Poisson Alert Stream (for Multiple Log Events)*

In normal state $\theta_o$, alerts arrive at $\lambda_o$ whereas in attack state $\theta_1$, alerts arrive at $\lambda_1 = \lambda_o + \lambda_A$ (where $\lambda_A$ is the adversary attack signature rate).

According to Poisson's distribution, $P((k|\lambda) = \frac{(\lambda\Delta t)^k e^{-\lambda\Delta t}}{k!}$, the k aggregated SIEM alerts received over observation window $\Delta t$ updates the continuous $\mu\,(t + \Delta t)$.

The standard Poisson probability mass function (PMF) for observing k events in a time window $\Delta t$ at rate λ is given by equation (11)

$$P(k|\lambda) = \frac{(\lambda\Delta t)^k e^{-\lambda\Delta t}}{k!} \qquad (11)$$

From the Bayes' Rule in equation (9) the updated Events per second (EPS) ($\mu(t + \Delta t) = P(\theta_1|k)$) is shown in equation (12)

$$\mu(t + \Delta t) = \frac{P(k|\theta_1).\mu(t)}{P(k|\theta_1).\mu(t) + P(k|\theta_0).(1-\mu(t))} \qquad (12)$$

Under an active attack ($\theta_1$) the events arrive at a rate of $\lambda_1$, the Poisson's probability is given by equation (13).

$$P(k|\theta_1) = \frac{(\lambda_1\Delta t)^k e^{-\Delta t\lambda_1}}{k!} \qquad (13)$$

Similarly, under normal noise ($\theta_0$) the events occur at a rate of $\lambda_0$.

$$P(k|\theta_0) = \frac{(\lambda_0\Delta t)^k e^{-\Delta t\lambda_0}}{k!} \qquad (14)$$

Substituting equation (13) and (14) in equation (12).

$$\mu(t + \Delta t) = \frac{(\lambda_1)^k e^{-\Delta t\lambda_1}.\mu(t)}{(\lambda_1)^k e^{-\Delta t\lambda_1}.\mu(t) + (\lambda_0)^k e^{-\Delta t\lambda_0}.(1-\mu(t))} \qquad (15)$$

*D3. Derivation of the Stackelberg trigger threshold for MGM Resorts Case*

On any given day, the baseline probability undergoing an active, high impact ransomware breach is statistically low at 1-2% and P(E) is 0.739 (calculated from equation 2) and $E[L_{FAIR}]$ is estimated at $92.58 million from Fig.2 (Monte Carlo simulation). Hence, the *Expected Unmitigated Risk* is calculated at $1.36 million using equation (16).

$$Expected\ Unmitigated\ Risk = \pi_0\ P(E)\ E[L_{FAIR}] \quad (16)$$

Since the *Expected Risk* is lower than immediate operational downtime cost of executing proactive containment ($C_D \approx$ \$15 *million*), a static system will always choose to do nothing (defender action- $d_l$).

The $C_D$ represents the cost of executing proactive containment (e.g. immediate session revocation, isolating active directory domains, localized service throttling). In the MGM Case Study, $C_D$ is approximated at $15 Million because according to the Caesar's Entertainment SEC filling and public disclosures, they chose to execute a controlled negotiation and containment strategy to avoid unmitigated catastrophic operational losses; that hit MGM Resorts and exceeded $100 million in response cost.

The critical threshold, $\mu^*$ represents the optimal point where the risk of inaction outweighs the cost of containment.

Mathematically, it is calculated using equation (17) at 0.219 or 21.9%.

$C_D = \mu^* \, [P(E) \, E[L_{FAIR}]\,]$ (17)

*D.4 Derivation of Events per second (EPS) and Stackelberg trigger threshold*

Assuming a normal operation baseline $\lambda_0 = 5000 \, EPS$ with an elevated log rate during active exploitations of $\lambda_A = 6200 \, EPS$ for an enterprise -wide raw SIEM baseline rate. Hence the corresponding events arrive at the rate as shown in

$\lambda_1 = \lambda_A + \lambda_0 = 6200 \, EPS$ (18)

The background arrival rate ($\lambda_0$) can be empirically adjusted using occupancy or device profiling to prevent false-alarm cascades in a high-density environment.

The prior probability of an active breach on any given day is low ($\pi_0 = 2\%$) and detection probability for high fidelity such as identity credential resets from an unrecognized device are $P_d \in [0.85, 0.95]$. Similarly, false positive or noise floor rate is low at $P_{fp} \in [0.01, 0.05]$.

*Case A: Processing no alerts (z= 0)*

If no alert is observed (z=0), $P(z = 0|\theta_1)$=0.10 and $P(z = 0|\theta_0) = 0.97$

$$\mu(z = 0) = \frac{(1-P_d).\pi_0}{(1-P_d).\pi_0 + (1-P_{fp}).(1-\pi_0)} \quad (19)$$

$$\mu(z = 0) = \frac{0.10 \ x \ 0.02}{(0.10 \ x \ 0.02) + (0.97 \ x \ 0.98)}$$

= 0.0021=0.21%

The absence of an alert suppresses the active attack belief down to 0.21% preserving uptime and avoiding unnecessary containment cost ($C_D$).

*Case B: Processing an alert (z=1)*

When a suspicious identity activity triggers an alert (z=1), the defender updates their posterior belief.

$$\mu(z = 1) = \frac{0.90 \ x \ 0.02}{(0.90 \ x \ 0.02) + (0.03 \ x \ 0.98)} \quad (20)$$

= 0.3793 =37.93%

When an alert vector z=1 is ingested, the posterior belief $\mu\,(z)$ updates dynamically from a low baseline prior $\pi_o$ to an actionable threat probability. In our parameterized case study of MGM Resorts, a single high fidelity identity alert should have elevated $\mu\,(z)$ from 2% to 37.9 %, instantly surpassing the optimal Stackelberg trigger threshold ($\mu^* = 0.219$) and initiating Automated Proactive Monitoring (d2, defense policy-2). This can also be attributed to forecasted monetary loss; the Expected Loss of doing nothing rises to \$26 million (=0.38*\$68.42 million) which is greater than $C_D$.

Fig.3 summarizes the Finite State transition model governing enterprise network posture under continuous Bayesian Filtering. In this formulation, operational states are classified into a Normal Baseline State ($\theta_0$) and a Suspicious Activity State ($\theta_1$). State transitions are driven by real time SIEM alert vectors (z $\in$ {0,1}) and evaluated by comparing the updated posterior belief $\mu(t)$ against the optimal Stackelberg decision threshold.

The system remains in nominal operation ($\theta_0$) under quiet telemetry (z =0), where $\mu(t) < \mu^*$. Ingestion of anomalous log vectors (z=1) transitions the belief state into the dynamic threat zone ($\theta_1$). If subsequent observations demonstrate quite operational trends (z=0), posterior updating decay restores the belief state to the normal baseline. However, if sustained anomalous telemetry escalates the posterior belief such that $\mu(t) \geq \mu^*$, the control engine automatically triggers automated containment ($d_1$), forcing a deterministic transition into Proactive Isolated State to halt adversary lateral movement.

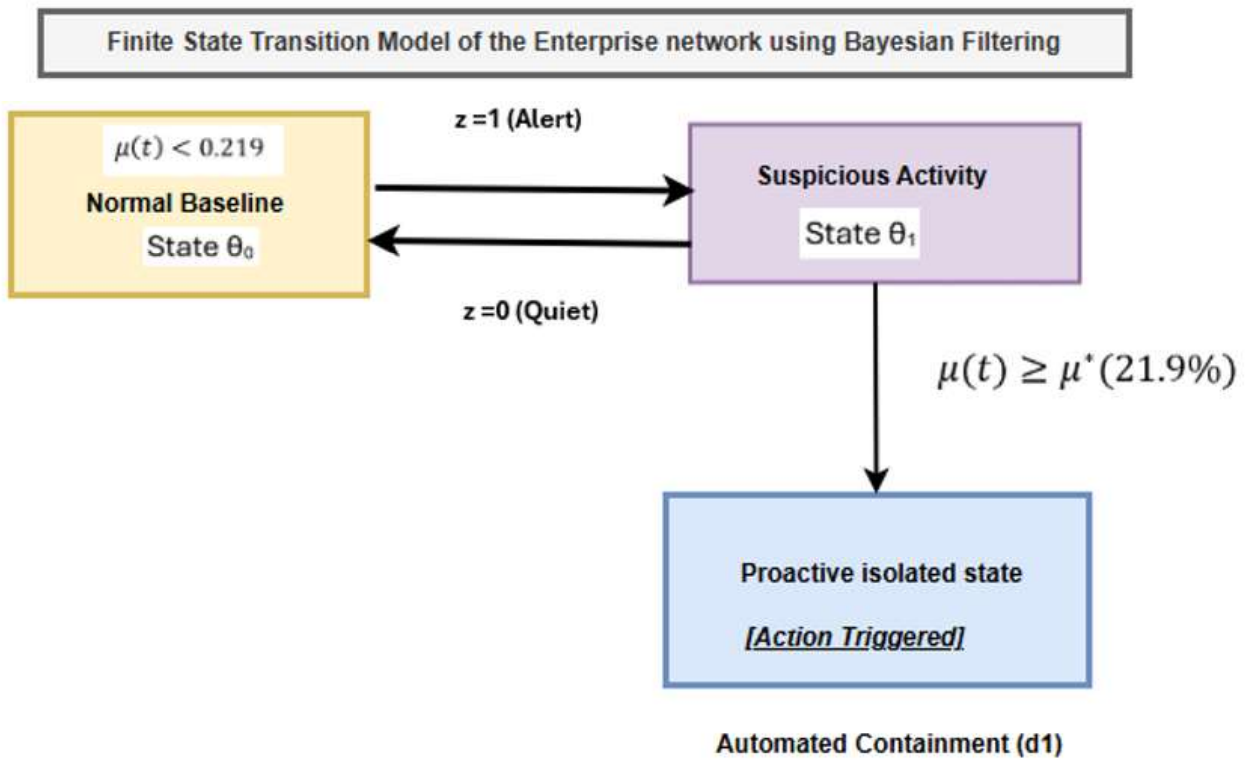

**Fig.3**: Finite State Transition model of the Enterprise network using Bayesian Filtering and Stackelberg control.

## IV. Strategic Game Formulation and Payoff Matrix

To evaluate optimal decision-making under state uncertainty-such as mitigating lateral movement during the MGM Resorts cyber breach scenario-the operational interaction is modeled as a non-cooperative Bayesian-Stackelberg game ($\Gamma$) between the *Enterprise Defender (D)* and the external *Adversary (A)*. The strategic game is defined by equation (21).

$\Gamma = <(N), \{S_D, S_A\}, \{U_D, U_A\}>$

(21)

where *N*= {*D*, *A*} represents the set of decision-making players, $S_D$ and $S_A$ denote the discrete strategy spaces available to the defender and adversary respectively. Similarly, *U*= $\{U_D, U_A\}$ defines their respective expected utility functions.

While foundational cybersecurity game-theoretic models reduce player interactions to binary outcomes-namely, (i) *No Attack/ No Defense*, and (ii) *Attack/Defense* [33,34]-more granular formulations, such as Attiah *et al*. [33], stratify operational intensity into multi-level activity tiers (for example, nominal baseline, low intensity probing and high intensity exploitation). In our formulation, player strategy spaces are structured to map directly to enterprise operational states:

$$S_D = \{d_1(Proactive\ Containment), d_2(Passive\ Monitoring)\}$$

$$S_A = \{a_1(Full\ Breach, Lateral\ Movement\ ), a_2(Abort, Probe\ Failure)\}$$

The optimum defender decision policy relies on evaluating the operational cost-to-benefit ratio against the dynamic threat telemetry. The Utility functions, $U_D$ and $U_A$ are parameterized using the empirical risk metrics:

$U_D(d_1, a_1) = -C_D$

$U_D(d_2, a_1) = -P(E).E[L_{FAIR}]$

The adversary's utility $U_A(d, a)$ represents their net payoff based on the target value gained $V_A$ from the operational expenditure or exploit investment cost, $C_A$ incurred during the attack.

$U_A(d_1, a_1) = -C_A$

$U_A(d_2, a_1) = P(E).(V_A - C_A)$

The $V_A$ is modeled as a fraction of the primary asset or extortion demand and for this case study can be simulated between \$15 million to \$30 million which is in line with Caesar's initial demand. The attacker's investment $(C_A)$ represents the initial access broker (IAB) credentials in purchasing, phishing, vishing infrastructure, tool development, operative overhead among others. Since $C_A \ll C_D$, in this study it can be simulated between \$50k to \$250k.

Table-II summarizes strategic game matrix across utility pairs $(U_D, U_A)$ comparing expected defender operational losses against adversary net payoff based on empirical calculations.

TABLE II

SUMMARY OF THE STRATEGIC GAME MATRIX ACROSS UTILITY PAIRS OF ENTERPRISE DEFENDER (D) AND THE EXTERNAL ADVERSARY (A)

| **Defender Strategy ($S_D$)** | **Adversary action: Breach/ Lateral movement ($a_1$)** | **Adversary action: Abort/ Probe failure ($a_2$)** |
|---|---|---|
| Proactive Containment | *Adversary:*$U_A(d_1, a_1) = -C_A$ | *Adversary* $U_A(d_1, a_2)=0$ |
| | *Defender:*$U_D(d_1, a_1) = -C_D$ | *Defender:* $U_D(d_1, a_2) = -C_D$ |
| Passive/ Reactive ($d_2$) | *Adversary:*$U_A(d_2, a_1) = P(E).(V_A - C_A)$ | *Adversary* $U_A(d_2, a_2)=0$ |
| | *Defender:*$U_D(d_1, a_1) = -P(E).E[L_{FAIR}]$ | *Defender:* $U_D(d_2, a_2)=0$ |

## I. Simulation Results and Discussion

To operationalize the game-theoretic optimization and recursive Bayesian updating, the step-by-step decision control logic is formalized in Algorithm 1. The algorithm continually ingests SIEM telemetry, computes the Poisson likelihood ratio and evaluates the posterior belief against the analytical Stackelberg decision boundary to execute automated isolation.

Fig. 4 illustrates the performance of the proposed dynamic Bayesian updating scheme and automated Game-Theoretic control execution across a simulated 100-second cyber breach scenario ($\Delta t = 1$). Assuming the operational baseline background noise was parameterized at $\lambda_0$= 50 EPS, while the elevated adversary activity signature was set at $\lambda_A$ =30 yielding an active attack ingestion rate of $\lambda_1 = 80EPS$ (as a filtered security event channel rate). The containment cost ($C_D$=\$15M), CVSS Exploitability Probability (P(E) ≈0.739) and Expected Loss ($E[L_{FAIR}] \approx 92.58M$) were directly calibrated from the MGM Resorts study yielding a critical Stackelberg decision threshold, $\mu^* = 0.219$ using equation (17).

To model adversary incursion, an active attack window was introduced between *t=35s* and *75s*, resulting in elevated telemetry ingestion (EPS > 50) during this interval. Starting from a baseline prior threat probability of $\pi_0$ =0.02, incoming log telemetry was stochastically generated using the MATLAB "*poissrnd*" function. The posterior belief-state $\mu(t)$ was dynamically updated for unobserved (z = 0) and observed (z =1) anomaly vectors using equation (19) and (20) respectively. As shown in Fig.4(B), $\mu(t) < \mu^*$, the defender maintains a *Passive Operational* posture. However, upon ingesting sustained anomalous telemetry during the attack window, the posterior belief rapidly escalates, crossing the critical threshold at t=58s ($\mu(t) \geq 0.219$) to execute *Automated Proactive Containment.*

**Algorithm 1:** The pseudocode for Dynamic Bayesian-Stackelberg Containment Loop

**Input:** CVSS Sub-scores $(AV, AC, PR, UI)$, FAIR Loss Distribution (Min,ML,Max), Prior Belief $\pi_0$, Baseline Noise Rate $\lambda_0$, Attack Signal Rate $\lambda_1$, Telemetry Window $\Delta t$.
**Output:** Containment Threshold $\mu^*$, Posterior Belief Trajectory $\mu(t)$, Automated Isolation Action $a^*$.

**Compute Exploitability Probability:** $P(E) = 8.22 \times AV \times AC \times PR \times UI$
**Compute Attacker Payoff:** $U_A = P(E) \cdot E[L_{\text{FAIR}}] - C_A$
**Compute Expected Risk:** Run PERT Monte Carlo sampling for $E[L_{\text{FAIR}}]$
**Set Optimal Stackelberg Threshold:** $\mu^* = \frac{C_D}{P(E) \cdot E[L_{\text{FAIR}}]}$ $(\mu^* \approx 0.219)$
**Initialize Belief State:** $\mu(0) \leftarrow \pi_0$
**For each** time step $t = 1,2,\ldots,N$ **do**
    Ingest telemetry signal $k_{\text{obs}}$ over window $\Delta t$ ($k_{\text{obs}} \sim \text{Poisson}(\lambda)$)
    **Compute Likelihood Ratio:** $\mathcal{L} = \left(\frac{\lambda_1}{\lambda_0}\right)^{k_{\text{obs}}} \cdot \exp(-(\lambda_1 - \lambda_0)\Delta t)$
    **Update Posterior Belief:** $\mu(t) = \frac{\mathcal{L} \cdot \mu(t-1)}{\mathcal{L} \cdot \mu(t-1) + (1 - \mu(k-1))}$
    **Apply Belief Floor:** $\mu(t) = \max(\mu(t), \mu_{\min}))$
    **If** $\mu(t) \geq \mu^*$ **then**
        Trigger automated network isolation ($a^* \leftarrow$ Isolate)
        **Break** (Terminate loop and execute containment at $t = 58$s)
    **Else**
        Maintain continuous monitoring ($a^* \leftarrow$ Observe)
    **End If**
**End For**

Fig. 5 compares the sensitivity analysis of the Bayesian posterior belief trajectories $\mu(t)$ and automated containment time across various CVSS exploitability probabilities P(E). By integrating real-time Poisson distributed SIEM event streams over discrete time windows $\Delta t$, the engine dynamically computes the Likelihood ratio ($\mathcal{L}$), updates the posterior threat belief $\mu(k)$, and evaluates the state against the closed form Stackelberg boundary ($\mu^*$). This structure ensures sub minute automated isolation upon detecting persistent adversarial probing while suppressing transient telemetry noise below the decision boundary.

Under lower exploitability (P(E) = 0.450), the Stackelberg decision boundary relaxes to $\mu^*$ =0.360, requiring higher telemetry confidence and delaying automated isolation to 20s. Conversely, under higher exploitability (P(E) = 0.920), the threshold tightens to $\mu^*$ =0.176, accelerating automated SOAR control response to t=15s upon attack onset. This dynamic shift confirms that the control engine automatically balances intervention costs against live perimeter vulnerability.

Apart from the CVSS Exploitability probability there are other factors that are core parameters and influence the Stackelberg threshold and subsequent response time. Table -III compares the multi-parameter sensitivity analysis of Stackelberg Threshold. The isolation intervention cost directly influences the magnitude of Stackelberg decision boundary. If the $C_D$ increases from $15M to $45M due to broader business intervention, the threshold increases from 0.219 to 0.657. The engine becomes more cautious, requiring near certain Bayesian telemetry prior to triggering shutdown. Similarly, the FAIR financial loss is inversely correlated to the threshold. This is because the higher potential risk lowers the threshold for action, hence proactive containment can be triggered at first sign of reconnaissance. While telemetry parameters, SIEM detection sensitivity ($P_d$) and false positive rates ($P_{fp}$) do not change the magnitude of $\mu^*$ but effects how fast the posterior belief $\mu$ reaches $\mu^*$. High fidelity detector rule accelerates posterior updating, driving $\mu \geq \mu^*$ in fewer time steps. On the other hand, high background noise dilutes belief accumulation and causes delaying containment.

## V. Limitations, Future Direction and Conclusion

While the proposed parameterization framework establishes an empirical bridge between vulnerability scoring, financial risk modeling, and game theoretic containment, several modeling assumptions define its current operational boundaries.

1. *Static Vulnerability Boundary (P (E) Boundary):* The CVSS v3.1 exploitability score assumes "upper bound ease of access", treating the adversary's initial success probability during social engineering as a static parameter (P (E)=0.739) rather than a dynamically fluctuating variable during active probing.
2. *Unimodal Financial Loss Distribution (FAIR Model):* The Monte Carlo FAIR simulation models operational losses using a unimodal PERT (beta) distribution bounded by explicit financial estimates (Minimum=$80M, Mode=$94M and Max=$120M). This assumption, while effective for parameterization baseline risk, catastrophic cyber events may exhibit heavy tailored distribution under extreme scenarios.
3. *Binary Network State and Alert State Space* $(\theta, z)$ *:*The Bayesian updating engine models network state transitions within a binary space:$\theta \in \{\theta_0, \theta_1\}$ (normal, active reconnaissance). Similarly, SIEM telemetry is abstracted into independent binary alert events ($z \in \{0,1\}$) governed by constant sensitivity ($P_d$) and false positive rates ($P_{fp}$).
4. *Adversarial Rationality in Stackelberg Game:* The strategic decision model operates on the assumption of a rational adversary maximizing expected utility ($U_A = V_A - C_A$). The attacker acts as a deterministic utility maximizing agent rather than a non-deterministic or irrational threat actor.
5. *Deterministic Defender Action Costs* ($C_D$): The automated containment threshold ($\mu^* = 0.219$) assumes defender costs ($C_D = \$15M$) are deterministic and fixed. Calibrated against the empirical settlement baseline observed in the Caesars entertainment incident.

TABLE III

MULTI-PARAMETER SENSITIVITY ANALYSIS OF STACKELBERG THRESHOLD

| **Parameter under test** | **Baseline value** | **Test shift** | **Impact on threshold** | **Dynamic response time** | **Operational interpretation** |
|---|---|---|---|---|---|
| Intervention cost ($C_D$) | $15M | $5M | 0.219 to 0.073 | 22 s | Low isolation cost allows aggressive early containment. |
| Intervention cost ($C_D$) | $15M | $35 M | 0.219 to 0.511 | 114 s | High downtime cost forces higher telemetry certainly prior to isolation. |
| Expected Loss $E[L_{FAIR}]$ | $92.58M | $200 M | 0.219 to 0.101 | 28s | Catastrophic exposure forces proactive, early isolation. |

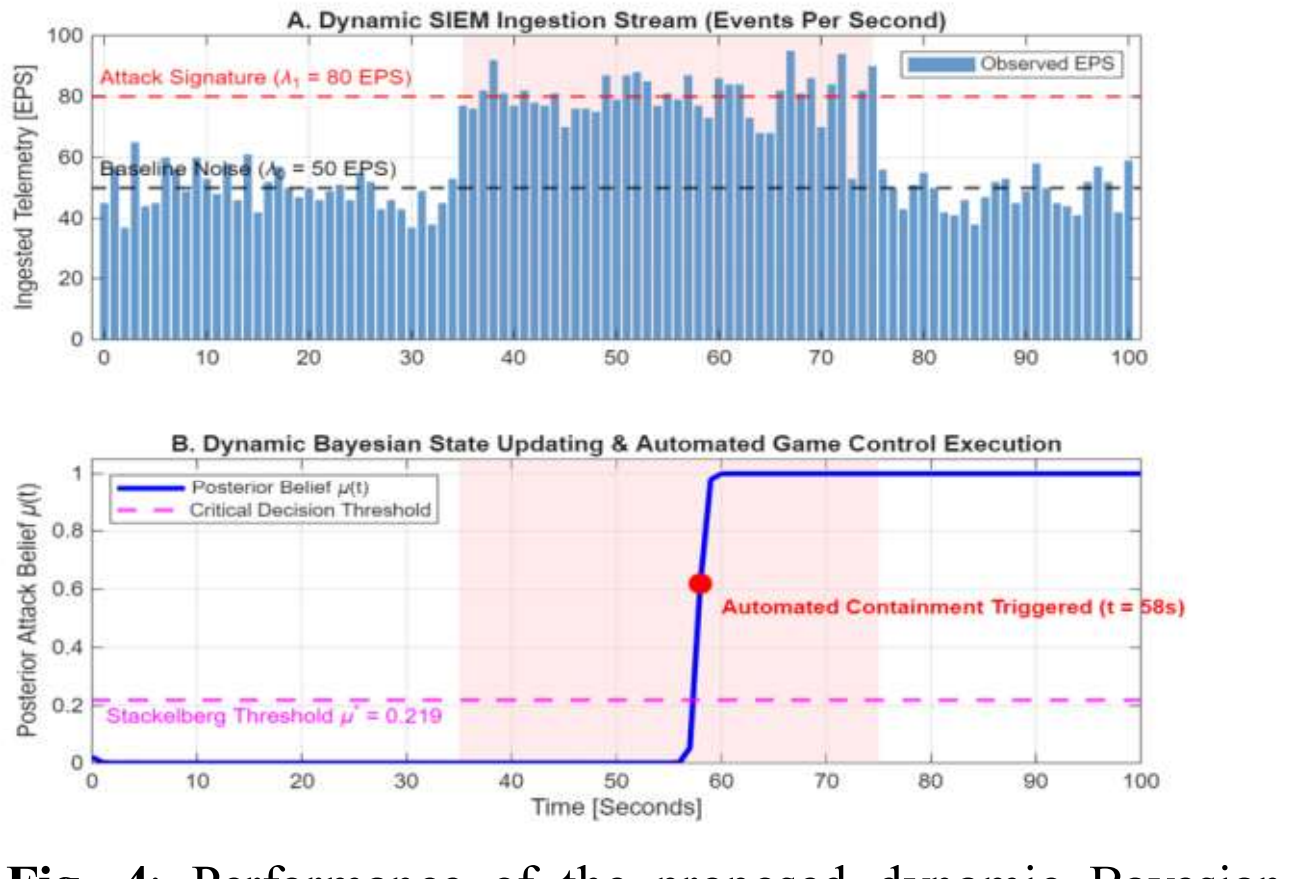

**Fig. 4**: Performance of the proposed dynamic Bayesian-

Stackelberg decision engine under fluctuating SIEM log arrival rates (EPS). (A) Observed telemetry stream highlighting background noise versus active adversary signatures. (B) Posterior attack probability trajectory triggering automated containment upon crossing critical Stackelberg decision threshold.

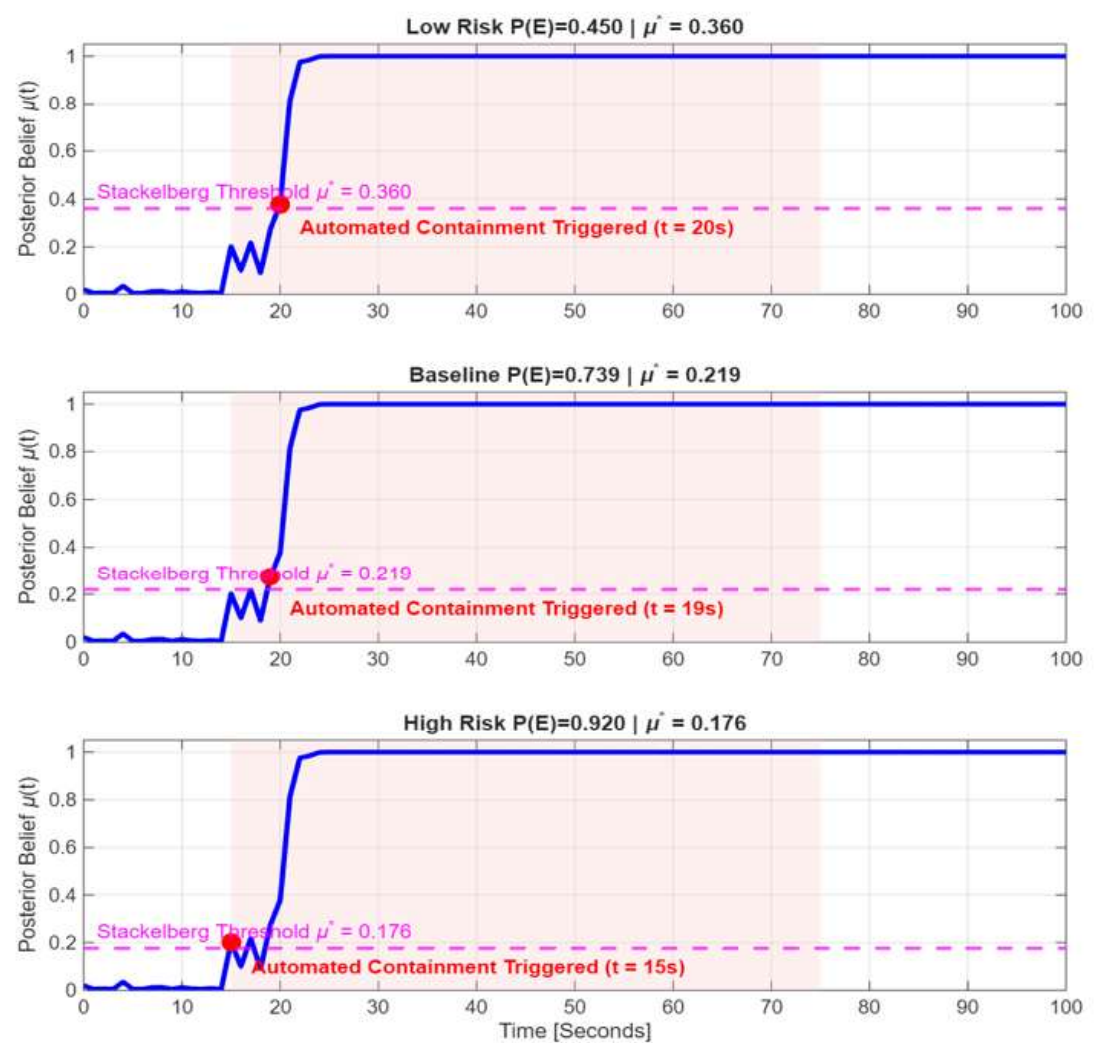

**Fig. 5:** Dynamic Stackelberg containment threshold shifts and SOAR trigger times under parametric variations of perimeter exploitability P(E).

To address these modeling limitations, future extensions of this work will focus on the following strategic avenues:

(i) *Nonstationary background telemetry and multi-state graphs:* Future work will expand the binary state space θ into continuous, dynamic Markov chain topologies to respect multi-stage kill chains. Additionally, integrating continuous time-windowed Poisson models and physical network baselines will enable adaptive filtering under non-stationary enterprise background traffic.

(ii) *Heavy tailed risk quantification*: Future iterations will evaluate generalized extreme values and Pareto Loss distributions within the FAIR framework to better capture high impact; low probability tail risks associated with zero-day exploits and multi-facility supply chain distribution.

(iii) *Automated Security Orchestration and Response (SOAR) and hardware enforcement*: Translating the Bayesian decision threshold into physical SOAR and hardware enforced isolation architectures will validate the framework's performance against live red-team exercises in active enterprise test systems.

In summary, this work advances game theoretic cyber defense beyond static, arbitrary payoff matrices by grounding defender utilities directly in CVSS v3.1 exploitability metrics and PERT-bounded FAIR stochastic financial loss distribution. An integrated pipeline is established that ingests real-time, Poisson distributed SIEM telemetry to evaluate network conditions between normal operational baseline ($\theta_o$) and active adversarial reconnaissance ($\theta_1$). By dynamically updating posterior threat beliefs $\mu$ (t) via Bayesian inference the framework evaluates the exact Stackelberg decision boundary relative to deterministic containment costs.

This yields a mathematically derived, closed form containment threshold that triggers automated network isolation upon detecting suspicious telemetry streams. Calibrated against empirical data from the 2023 MGM Resorts and Caesars Entertainment incidents, this framework proves that sub-minute automated isolation significantly restricts lateral movement and mitigates multi-million-dollar operational disruption compared to delayed manual response. Finally, this model provides SOC architectures with a reproducible control theoretic foundation for automating SOAR process prior to adversarial exploitation.

**Conflict of Interest**

The authors have no conflict of interest.

**Data Availability**

**GitHub Repository for the MATLAB Codes:**

https://github.com/Shadeeb-Hossain/Quantitative-Risk-Parameterization-Cybersecurity-

**Hugging Face Repository for the MATLAB codes:**

https://huggingface.co/datasets/shadeebhossain/Quantitative-Risk-Parameterization-Cybersecurity

**Funding Declaration**

No funding was received for this study.

.

**Shadeeb Hossain** received his PhD in Electrical Engineering from The University of Texas at San Antonio in May 2023 and his M.S. in Engineering from Central Michigan University. He is the Founder and Principal Engineer at the Research Division of Shadeeb Engineering Lab and serves as an Adjunct Faculty member at Capitol Technology University.
Dr. Hossain was a Research Fellow at Weill Cornell Medicine (2022-2023) and the recipient of the Materials Research Fellowship and G-TAP Award during his doctoral studies at The University of Texas at San Antonio.